\documentclass[twocolumn,superscriptaddress,aps,longbibliography,showpacs,preprintnumbers,amsmath,amssymb,floatfix]{revtex4-1}
\usepackage{titlesec}
\usepackage{amsmath}
\usepackage{bm}
\usepackage{graphicx}
\usepackage[ansinew]{inputenc}
\usepackage{array}
\usepackage{color}
\usepackage{amsxtra}
\usepackage{amstext}
\usepackage{amssymb}
\usepackage{latexsym}
\usepackage{gensymb}
\usepackage{dsfont}
\usepackage{braket}
\usepackage[caption=false]{subfig}
\usepackage[makeroom]{cancel}

\usepackage{balance}

\usepackage{dcolumn}
\usepackage{bm}
\usepackage{bbm}
\usepackage{braket}
\usepackage{mathrsfs}
\usepackage{amstext}
\usepackage{euscript}
\usepackage{txfonts}

\usepackage[unicode=true,
 bookmarks=false,
 breaklinks=false,pdfborder={0 0 1},backref=false,colorlinks=true,citecolor=blue]
 {hyperref}

\makeatletter
\@ifundefined{textcolor}{}
{%
 \definecolor{BLACK}{gray}{0}
 \definecolor{WHITE}{gray}{1}
 \definecolor{RED}{rgb}{1,0,0}
 \definecolor{GREEN}{rgb}{0,1,0}
 \definecolor{BLUE}{rgb}{0,0,1}
 \definecolor{CYAN}{cmyk}{1,0,0,0}
 \definecolor{MAGENTA}{cmyk}{0,1,0,0}
 \definecolor{YELLOW}{cmyk}{0,0,1,0}
}

\makeatletter
\renewcommand*\env@matrix[1][*\c@MaxMatrixCols c]{%
  \hskip -\arraycolsep
  \let\@ifnextchar\new@ifnextchar
  \array{#1}}
\makeatother

\newcommand{\cref}[1]{Ref.\,\cite{#1}}

\makeatother

\begin{document}

\title{Stereographic Geometry of Coherence and Which-path Information}

\author{Yusef Maleki}
\email{maleki@physics.tamu.edu}
\affiliation{Department of Physics and Astronomy, Texas A\&M University, 
	College Station, Texas 77843-4242}

\date{\today}

\begin{abstract}

Recently, it was shown that quantum entanglement is an indispensable part of the duality behavior of  light. Here, we report a surprisingly intimate connection between the stereographic projection and the duality--entanglement nature of a single photon. We show that, the duality--entanglement relation [Optica \textbf{5}, 942
(2018)], naturally
emerges from the stereographic projection geometry.
We demonstrate that this geometry is complementarity
sensitive; in the sense that, it is sensitive to the particle
nature, wave nature, and entanglement nature of a
single photon.
\end{abstract}

\maketitle

The attribution of wave nature to particles by de Broglie was one of the most profound counterintuitive weirdnesses of  quantum physics \cite{de1923waves}. Later, this bizarre nature of a single quantum was introduced by  Bohr in a more general setting,  as the complementarity principle; where, a quantum object possesses properties which are equally real but mutually exclusive \cite{bohr1928quantum}. 

The duality nature of a single quantum becomes manifest by considering a photon subjected to an interferometer that can demonstrate wave or particle properties \cite{scully1991quantum}. The particleness is determined by the amount of information one  attains on the path distingishablity of photon \cite{scully1991quantum,durr1998origin}. Alternatively, the waveness can be characterized by the visibility of the interference pattern on the screen \cite{scully1991quantum,durr1998origin}.

Richard Feynman considers the wave-particle duality incorporating the underlying mystery of quantum mechanics. More precisely, he says: "In reality, it contains the only mystery" \cite{feynman1965lectures}. 

The 'mystery' pointed to by Feynman, remains as a long--standing  subject of  dedicated investigations \cite{jacques2008delayed,chang2008quantum,arndt1999wave,rab2017entanglement,menzel2012wave,peruzzo2012quantum}, providing an underlying platform for most of the interesting features and far--reaching application of quantum mechanics \cite{jacques2007experimental,kim2000delayed,kaiser2012entanglement,jeong2013experimental,bagan2016relations,winter2016operational,cheng2015complementarity,lostaglio2015quantum,baumgratz2014quantifying}. However, despite dedicated efforts, the quantitive description of the wave-particle duality was not found until 1979;  where, Wootters and Zurek quantified  the wave-particle duality nature of a single quantum \cite{wootters1979complementarity}. Later, the duality nature of a quantum system was explicitly formulated as an inequality, where the visibility of the interference pattern $V$   and the which--path distinguishability $D$  of a single quantum, in a two-dimensional Hilbert space, was shown to satisfy $V^2+D^2\leq 1$ \cite{wootters1979complementarity,englert1996fringe}.

Recently, it was shown that  analysis of the vector mode coherence in Young double slit experiment uncovers the significant role of the entanglement on the duality nature of the light modes \cite{qian2018entanglement,qian2016coherence}. 
Accordingly, for the double slit experiment, the duality relation can be expressed through  $V^2+D^2+C^2= 1$  \cite{qian2018entanglement,qian2016coherence}. The concurrence, $C$, quantifies the entanglement of the two vector modes involved in the double slit experiment.

On the other hand, a pure two--dimensional quantum state  could be expressed via complex variables $\alpha$ and $\beta$ as $|\psi\rangle=\alpha  |0\rangle+\beta |1\rangle $, enabling the state to be represented as a point on  the Bloch sphere. The Bloch sphere can be obtained
from Hopf fibraion of $S^3$ \cite{lee2002qubit}. In which,  $S^3$  is the unit sphere embedded in $\mathbb{R}^4$,  the fibre $S^1$ is the global $U(1)$ phase of the state and $S^2$  the base manifold.  Similarely, two--qubit states, are characterized by $S^7$ sphere embedded in $\mathbb{R}^8$, providing  fibration geometry  in the quaternionic skew-field \cite{mosseri2001geometry,najarbashi2007geometry}. In this platform, the product states are mapped into a 2-dimensional planar subspace, enabling  entangled and separable subspaces be distiguishable through this map \cite{mosseri2001geometry}.

In this lettter, we show that the duality--entanglement relation $V^2+D^2+C^2= 1$, naturally emerges from the stereographic projection \cite{lee2002qubit,najarbashi2007geometry} obtained from fibration of the $S^7$ sphere geometry. In this setting, there is no need to specify the experimental setup, as the emergence of the relation is a generic property of the geometry.
The full geometry of the duality is captured by $S^4$ sphere, obtained from Hopf fibration of $S^7$. This enables stratification of the duality relation through entanglement, providing a full geometric picture of the scenario. This surprisingly intimate connection between the  $S^7$  Hopf fibration  and the complementarity nature of a single photon is shown to be complementarity sensitive; in the sense that, it is sensitive to the particleness, waveness, and entangledness  of a single photon. 


We consider a generic scenario, where a single photon could be correlated with some other  system (be it an atom, photon, environment, etc). In this case , we assume the two-state photon Hibert space to be encoded by the basis $ |0\rangle$ and $ |1\rangle$ (note that these basis can be two orthonormal polarization degrees of freedom, photon number states, etc). Thus,
the generic form of such a state can be given as 
\begin{equation}\label{}
|\psi\rangle =\mu |0\rangle |\chi_1\rangle  +  \nu  |1\rangle  |\chi_2\rangle,
\end{equation}
where $ |\chi_1\rangle$ and $ |\chi_2\rangle$ are associated to any correlated system. We can write $ |\chi_1\rangle=\sum a_{i.j,...k}  |a_i \rangle |b_j \rangle...|c_k \rangle$, incorporating any involved degrees of freedom. Similarly, $|\chi_2\rangle=\sum b_{i.j,...k}  |a_i \rangle |b_j \rangle...|c_k \rangle$.
Where, $ |\chi_1\rangle$ and $ |\chi_2\rangle$  are not necessarily orthogonal; however, they can always be considered to span a two dimentsional vector space \cite{maleki2019linear,maleki2018witnessing,maleki2018entangled}. Thus, they can always be mapped into a two dimensional  space such that 
$ |\chi_1\rangle=a |e\rangle+b  |f\rangle$  and $ |\chi_2\rangle=c|e\rangle+d  |f\rangle$ \cite{maleki2019linear,maleki2018witnessing,maleki2018entangled}. Here, $|e\rangle$ and $|f\rangle$ form orthonormal basis of the two dimensional Hilbert space. The coefficients are  determined by decomposition of the vectors $ |\chi_1\rangle$ and $ |\chi_2\rangle$ in the   $|e\rangle$ and $|f\rangle$ directions \cite{maleki2019linear,maleki2018witnessing,maleki2018entangled}. Thus, the most general form of the scenario can be written as 
\begin{equation}\label{generic}
|\psi\rangle = \alpha_0  |0\rangle |e\rangle  + \alpha_1  |0\rangle  |f\rangle  + \alpha_2  |1\rangle  |e\rangle + \alpha_3  |1\rangle  |f\rangle.
\end{equation}

In the two dimensional Hilbert space framework of the photon spanned by the bases $\lbrace |0\rangle, |1\rangle \rbrace$, one can assign Pauli matrices $\sigma_x= |0\rangle\langle 1|+|1\rangle\langle 0|$, $\sigma_y= -i|0\rangle\langle 1|+i|1\rangle\langle 0|$ and $\sigma_z= |0\rangle\langle 0|-|1\rangle\langle 1|$  to describe the system. 
 The duality nature of a correlated photon reduces to considering the state in Eq. \ref{generic}. The wave nature of the photon is determined by the fring visibility \cite{qian2019quantum}
 \begin{equation}
V= \frac{p_{D}^{max}-p_{D}^{min}}{p_{D}^{max}+p_{D}^{min}}.
\end{equation}
 To determine the visibility we can measure the observable $\frac{1}{2}(1+\sigma_x)$ such that $
 p_{D}= \langle \psi | \frac{1}{2}(1+\sigma_x)|\psi\rangle$ \cite{englert1996fringe}. This results in $
  p_{D}=\frac{1}{2}(1+2|\gamma| \cos\varphi )$,  where $\gamma=|\gamma| e^{i\varphi}=\bar{\alpha}_2\alpha_0 + \bar{\alpha}_3\alpha_1$. Here,  $\bar{\alpha}_i$ is the complex conjugate of ${\alpha}_i$.
Hence, the visibility of a single photon is given by 
\begin{equation}\label{V}
V=2|\bar{\alpha}_2\alpha_0 + \bar{\alpha}_3\alpha_1 |.
\end{equation}
On the other hand, the particle nature of the photon is related to our \textit{priori} knowledge on 
the predictability of the photon being in the state $ |0\rangle$ or $ |1\rangle$. Therefore, the particle nature of the photon can be quantified as \cite{qian2018entanglement,qian2016coherence}
\begin{equation}
D= \frac{|p_{0}-p_{1}|}{p_{0}+p_{1}},
\end{equation}
where, $p_{0}$ and $p_{1}$ are the probabilities of the photons being detected in the states $ |0\rangle$ and $ |1\rangle$, respectively.

From Eq. \ref{generic} we have, $p_{0}= |\alpha_0|^2+|\alpha_1|^2$ and $p_{1}= |\alpha_2|^2+|\alpha_3|^2$. Thus, the particleness  is given by
\begin{equation}\label{D}
D= |(|\alpha_0|^2+|\alpha_1|^2)- (|\alpha_2|^2+|\alpha_3|^2)|.
\end{equation}

The  relations above can be understood by considering  the reduced density matrix of the photon 
\begin{align*}
\rho_{ph}=
&\left(
  \begin{array}{cc}
|\alpha_0|^2+|\alpha_1|^2& (\bar{\alpha}_2\alpha_0 + \bar{\alpha}_3\alpha_1 )\\
 ({\alpha}_2\bar{\alpha}_0 + {\alpha}_3\bar{\alpha}_1 ) &|\alpha_2|^2+|\alpha_3|^2 \\       
  \end{array}
\right).
 \end{align*}
Accordingly, the waveness is determined by the coherence terms of the density matrix and the particlness is determined by our knowledge on the probabilities of finding the system in each basis, as expected.

It is remarkable that, since Eq.\ref{generic} is expressed as a two-qubit superposition, one may consider the photon state to be encoded in the second subsystem with the bases $\lbrace  |e\rangle, |f\rangle  \rbrace$. In this scenario, $V$ and $D$ can be determined through the reduced density matrix of the second subsystem; where, $V=2|\bar{\alpha}_1\alpha_0 + \bar{\alpha}_2\alpha_1 |$, and $D= |(|\alpha_0|^2+|\alpha_2|^2)- (|\alpha_1|^2+|\alpha_3|^2)|$.

Now, to develop a stereographic projection platform for the complementarity concept of the photon, we introduce the quaternification  map $\mathcal{F}$, such that it maps any $|\psi\rangle \in H^\mathbb{C}_2\otimes H^\mathbb{C}_2$ to $|\psi\rangle_q \in H^\mathbb{Q}_2$ \cite{najarbashi2007geometry,najarbashi2017quantum}
\begin{equation}
\mathcal{F}(|\psi\rangle):=|\psi\rangle_q= \ q_1  |0\rangle_q + \ q_2  |1\rangle_q =
\left(
  \begin{array}{c}
q_1\\
q_2\\       
  \end{array}
\right).
\end{equation}
Where, $ |0\rangle_q$ and  $|1\rangle_q$ are the basis of spinors of the quaternionic  states. $q_1$ and $q_2$ are  quaternion numbers defined as
\begin{equation}\label{}
q_1= \alpha_0 + \alpha_1 \hat{e}_2, \qquad  q_2 = \alpha_2 + \alpha_3 \hat{e}_2 ,
\end{equation}
 satisfying the normalization condition $|q_1|^2+|q_2|^2=1$.

Recall that, quaternions are usually presented by four real numbers $x_0,x_1,x_2,x_3$, such that $
q=x_0 \hat{e}_0+x_1 \hat{e}_1+x_2 \hat{e}_2+x_3 \hat{e}_3,$
with $\hat{e}_0=1$, and $\hat{e}_1^2=\hat{e}_2^2=\hat{e}_3^2=\hat{e}_1\hat{e}_2\hat{e}_3=-1$  as the imaginary units of the quaternionic algebra. Therefore, defining the complex numbers $z_1=x_0 \hat{e}_0+x_1 \hat{e}_1$ and $z_2=x_2 \hat{e}_0+x_3 \hat{e}_1$, one has $q=z_1+z_2 \hat{e}_2$.

 The quaternionic stereographic projection $\mathcal{P}$, maps the points on $S^7$ to the extended quaternionic space $\mathbb{\tilde{Q}}\equiv\mathbb{Q} \cup \lbrace\infty\rbrace$  through \cite{mosseri2001geometry,najarbashi2007geometry}
\begin{equation}\label{P}
\mathcal{P} : (q_1,q_2)  \rightarrow  Q={q_1 q_2^{-1}}=\dfrac{1}{|q_2|^2}(\pi_1+\pi_2  \hat{e}_2),
\end{equation}
Where, $\pi_1=(\bar{\alpha}_2\alpha_0 + \bar{\alpha}_3\alpha_1 )$ and $\pi_2=(\alpha_1 \alpha_2-\alpha_0 \alpha_3).$

Interestingly, comparing with Eq. \ref{V}, $\pi_1$ is associated to the coherence term and the visibility can be given by $V=2|\pi_1|$. Moreover, the term $\pi_2$ is related to the concurrence measure of the two--qubit entanglement \cite{wootters1998entanglement} through $C=2|\pi_2|$. The concurrence measure of the state \eqref{generic} is defined through $C=|\langle \psi| J (\sigma_y\otimes \sigma_y)|\psi \rangle|$, where the operator J is an antilinear operator, such that $\langle \psi| J=\langle \psi^*|$ \cite{wootters1998entanglement}.

The second map is an inverse stereographic projection from $\mathbb{\tilde{Q}}$  onto the unit sphere $S^4$ defined by \cite{najarbashi2007geometry,mosseri2001geometry}
\begin{equation}\label{Q}
\mathcal{Q}: Q \rightarrow \{ \lbrace x_i  \rbrace, i=0,1,2,3,4 \}\  \text{with}  \ \sum^{4}_{i=0} \ x^2_i =1,
\end{equation}
where, the coordinates  $x_i$s are given by 
\begin{align}\label{x_0}
	x_0&=\langle \psi|  (\sigma_z\otimes I_a)|\psi \rangle=|q_1|^2-|q_2|^2, 
	\\
	\label{x_1}
	x_1&=\langle \psi|  (\sigma_x\otimes I_a)|\psi \rangle=2 \text{Re}(\pi_1),
	\\
	\label{x_2}
	x_2&=\langle \psi|  (\sigma_y\otimes I_a)|\psi \rangle=2 \text{Im}(\pi_1),
	\\
	\label{x_3}
	x_3&=\text{Im}[\langle \psi| J (\sigma_y\otimes \sigma_y)|\psi \rangle]=2\text{Re}(\pi_2),
	\\
	\label{x_4}
	x_4&=\text{Re}[\langle \psi| J (\sigma_y\otimes \sigma_y)|\psi \rangle]=2 \text{Im}(\pi_2).
\end{align}
Surprisingly, the coordinates of the stereographic conformal mapping provide
\begin{align}
\label{D2}
	D^2&= x_0^2,
	\\
	\label{V2}
	V^2&= x_1^2+x_2^2,
	\\
	\label{C2}
	C^2&= x_3^2+x_4^2.
\end{align}
Since, cooerdinates of the  $S^4$ hypersphere satisfy $\sum^{4}_{i=0} \ x^2_i =1$,  the relation $D^2+V^2+C^2=1$ emerges naturally. This provides a fully geomerical proof for the duality--entanglement relation based on the Hopf fibration of $S^7$. The analysis reveals that the duality--entanglement relation is a characteristics of the state-vectors geometry in the Hilbert space, which is independant of the platform by which  two-qubit state is physically realized. Therefore, the relation is valid beyond the scope of the vector mode coherence in Young double slit experiment \cite{qian2018entanglement}, and holds for any two--qubit pure quntum state.

According to Eq. \ref{P}, product states are mapped into the pure complex subspace in the quaternion field though the quaternionic stereographic projection $\mathcal{P}$. In other words, this map provides a one to one correspondence between the  points on complex plane $\mathbb{\tilde{C}}$ and separable single photon states. Also, when the wave nature of the photon vanishes ($\pi_1=0$), 
$\mathcal{P}$ reduces to subset of the quaterinionic field that is only spanned by the $\hat{e}_2$ and $\hat{e}_3$ coordinates of the system, such that 
$$
Q=\dfrac{1}{|q_2|^2}(\text{Re}(\pi_2)\hat{e}_2+ \text{Im}(\pi_2) \hat{e}_3)=\dfrac{1}{2|q_2|^2}(x_3\hat{e}_2+ x_4 \hat{e}_3).
$$
Similarly, states with no particle nature ($D=0$) are mapped into the unite quaternion $|Q|=1$ subspace of $\mathbb{\tilde{Q}}$.
States with only wave nature ($V=1$) are mapped into the unite circle in the subspace spanned by $\{ \hat{e}_0,\hat{e}_1\}$, and the ones with only particle nature ($D=1$) are mapped onto the $Q=\{0, \infty \}$ subspace of $\mathbb{\tilde{Q}}$. Finally, states with only entangled nature ($C=1$) are mapped into  the unite circle in the subspace spanned by $\{ \hat{e}_2,\hat{e}_3\}$, and product state are never mapped to this subspace.  
This proves that Hopf fibration is entangledness, waveness and particlness sensitive.


Separable states of the photon are mapped into $\mathbb{\tilde{C}}$. Geometrically, this means that $S^7$ is reduced to the two-dimensional planer subspace of $\mathbb{\tilde{Q}}$, i.e.,  $S^2$ sphere described by $x^2_0+x^2_1+x^2_2=1$. From Eqs. \ref{x_0}--\ref{x_4} it is clear that on the base sphere $S^4$ of the Hopf map, only $x_3$ and $x_4$ components depend on the observable of the second subsystem of Eq. \ref{generic}. Thus, the second qubit lives on the $S^3$ fiber of the geometry.  This analysis suggests that for a pure photon on the  $S^2$ sphere we always have $V^2+D^2=1$. To consider this from a different perspective, we note that  the pure state of a photon can  be expressed as
\begin{equation}\label{pure}
|\psi\rangle = \cos(\theta/2)  |0\rangle +e^{i \varphi} \sin(\theta/2)   |1\rangle.
\end{equation}
The waveness of the photon in this scenario is given  by $V=2 \sin(\theta/2)\cos(\theta/2)$ and the particleness by $D=|\cos(\theta/2)^2-\sin(\theta/2)^2|$. This immediately gives $V^2+D^2=1$.
For the mixed  photon state with the density matrix $\rho_{ph}$ we can obtain \cite{qian2019quantum}
\begin{equation}\label{}
V^2+D^2 =2 Tr(\rho_{ph}^2)-1,
\end{equation}
where the right--hand--side quantifies degree of pureness of the photon. For a maximally mixed state  $Tr(\rho^2)=1/2$, one has $V^2+D^2 =0$.

\begin{figure}
	\includegraphics[width=\columnwidth]{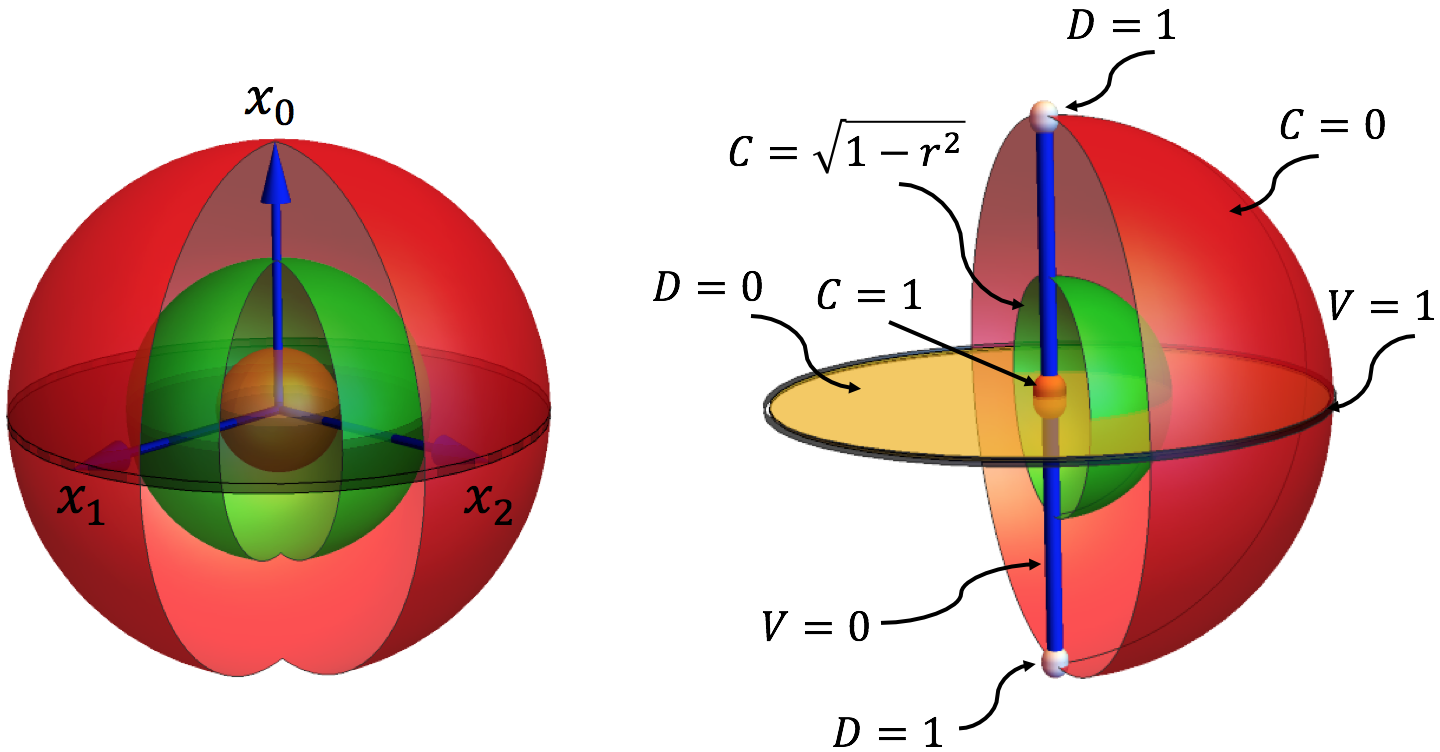}
	\caption{
 The geometry of the unit ball $B^3$, sliced as sells of radius $\sqrt{1-C^2}$. The boundary of the ball gives the $S^2$ sphere of the pure states, shown as the red (largest) shell. The inner shells represent entangled photon states, and the smaller shells correspond to the  larger entanglement. The center of the ball gives the maximally entangled states. The north and the south poles are associated with the pure particle nature of the photon with $D=1$. The points on the $x_0$ axis (shown in blue in the left panel) are the states with no wave nature. The great disc on $x_0=0$  corresponds to the states with not particle natures; where the boundary of the disc gives the great circle of the geometry corresponding to states having only the wave nature.
	}
	\label{fig:1}
\end{figure}

For the maximally entangled states, we have  $x_0=x_1=x_2=0$, where, both particleness and waveness nature of the photon vanishes, and the photon is left with the only entangledness. These states are given by the circle $x^2_3+x^2_4=1$ on the $S^4$ sphere, 
parametrized by real and complex parts of $2\pi_2$.

The relation $V^2+D^2 \leq 1$ can be understood by expressing it through Eqs. \ref{D}--\ref{V} as $x^2_0+x^2_1+x^2_2 \leq 1$. This demonstrates a unit ball $B^3$ of the radius 1.
From $V^2+D^2=1-C^2$, separable states cover the $S^2$ boundary of the ball (the Bloch sphere of the photon). States with the same entanglement can be represented in concentric spherical shells of radius $\sqrt{1-C^2}$ around the center, determined through $x^2_0+x^2_1+x^2_2=1-C^2$, with the maximally entangled states in the center of the ball (see Fig.1).

The north pole ($x_0=1$) is given by $|\psi\rangle_{Q=\infty} =  |0\rangle_{ph}  \otimes  (\alpha_0 |g_0\rangle+\alpha_1|e\rangle)$, where both $C$ and $V$ vanish; and, the only surviving part of the duality--entanglement scenario is the particleness with $D=1$. States with no quantum coherence are mapped into the points on the $x_0$ axis (see Fig.1). By decreasing $D$ along  $x_0$,
we attain states with entanglement determined via $1-x_0^2$. On this axis,  state \ref{generic} can be expressed in the Schmidt form
$|q_1| |0\rangle |\tilde{e}\rangle+|q_2| |1\rangle |\tilde{f}\rangle$, where $|\tilde{e}\rangle$ and $|\tilde{f}\rangle$ are two orthonormal basis for second subsystem. When the particleness becomes zero, the geometry reduces to the great disc of the ball at  $x_0=0$. The boundary of the disc is the circle $x^2_1+x^2_2=1$, where the photon has no entanglement or the particleness; rather, it only has the wave nature.


In conclusion, coherence and which-path information duality, serving as one of the most fundamental characteristics of the quantum mechanics, has remained as a long--standing subject of continued investigations. 
In this work, we have reported a surprisingly intimate correspondence between the Hopf fibration geometry of $S^7$  and the duality nature of a single quantum.  It was shown that this geometry is complementarity sensitive; in the sense that, it is sensitive to the particleness, waveness, and entangledness nature of a single photon. 

Even though we have considered a photon which could be correlated with some other systems, the results presented in this letter is valid for any pure two-qubit quantum state living in the Hilbert space $\mathcal{H}_2\otimes \mathcal{H}_2$.

\bibliography{ref}
\end{document}